\documentclass[aps,prd,reprint,nofootinbib,amsmath,amssymb,superscriptaddress]{revtex4-1} 
\usepackage{graphicx}   
\usepackage[
colorlinks=true,        
citecolor=blue,         
linkcolor=blue,         
urlcolor=blue           
]{hyperref}  
\usepackage{tabulary}
\usepackage{color}      
\usepackage{orcidlink}
\usepackage{units}

\newcommand{\nc}{\newcommand*} 
\newcommand{\beq}{\begin{equation}}
\newcommand{\eeq}{\end{equation}}
\def\({\left(}
\def\){\right)}
\def\[{\left[}
\def\]{\right]}
\nc{\Eq}[1]{Eq.~\eqref{#1}}     
\nc{\Fig}[1]{Fig.~\ref{#1}}     
\nc{\Table}[1]{Table~\ref{#1}}  
\nc{\Sec}[1]{Sec.~\ref{#1}}     
\nc{\red}[1]{\textcolor{red}{#1}}

\begin{document}
\title{Searching for Anisotropy in the Gravitational Wave Background Using the Parkes Pulsar Timing Array}

\author{Yiqin Chen\orcidlink{0009-0009-0136-7959}}
\affiliation{School of Physics and Astronomy, Beijing Normal University, Beijing 100875, China}
\affiliation{Department of Physics, Faculty of Arts and Sciences, Beijing Normal University, Zhuhai 519087, China}

\author{Shi-Yi Zhao\orcidlink{0009-0001-8885-5059}}
\affiliation{School of Physics and Astronomy, Beijing Normal University, Beijing 100875, China}
\affiliation{Department of Physics, Faculty of Arts and Sciences, Beijing Normal University, Zhuhai 519087, China}

\author{Zhi-Zhang Peng\orcidlink{0000-0001-9857-5504}}
\affiliation{School of Physics and Astronomy, Beijing Normal University, Beijing 100875, China}
\affiliation{Department of Physics, Faculty of Arts and Sciences, Beijing Normal University, Zhuhai 519087, China}

\author{Xingjiang Zhu\orcidlink{0000-0001-7049-6468}}
\email{zhuxj@bnu.edu.cn}
\affiliation{Department of Physics, Faculty of Arts and Sciences, Beijing Normal University, Zhuhai 519087, China}
\affiliation{Institute for Frontier in Astronomy and Astrophysics, Beijing Normal University, Beijing 102206, China}

\author{N. D. Ramesh Bhat\orcidlink{0000-0002-8383-5059}}
\affiliation{International Centre for Radio Astronomy Research, Curtin University, Bentley, WA 6102, Australia}

\author{Zu-Cheng Chen\orcidlink{0000-0001-7016-9934}}
\affiliation{Department of Physics and Synergetic Innovation Center for Quantum Effects and Applications, Hunan Normal University, Changsha, Hunan 410081, China}
\affiliation{Institute of Interdisciplinary Studies, Hunan Normal University, Changsha, Hunan 410081, China}

\author{Ma\l{}gorzata Cury\l{}o\orcidlink{0000-0002-7031-4828}}
\affiliation{School of Physics and Astronomy, Monash University, Clayton VIC 3800, Australia}
\affiliation{ARC Centre for Excellence for Gravitational Wave Discovery (OzGrav), Hawthorn, VIC, 3122, Australia} 

\author{Valentina Di Marco\orcidlink{0000-0003-3432-0494}}
\affiliation{School of Physics, University of Melbourne, Parkville, VIC 3010, Australia}
\affiliation{ARC Centre for Excellence for Gravitational Wave Discovery (OzGrav), Hawthorn, VIC, 3122, Australia} 

\author{George Hobbs\orcidlink{0000-0003-1502-100X}}
\affiliation{Australia Telescope National Facility, CSIRO, Space \& Astronomy, PO Box 76, Epping, 1710, NSW, Australia}

\author{Agastya Kapur\orcidlink{0009-0001-5071-0962}}
\affiliation{Australia Telescope National Facility, CSIRO, Space \& Astronomy, PO Box 76, Epping, 1710, NSW, Australia}
\affiliation{Department of Mathematics and Physical Sciences, Macquarie University, NSW 2109, Australia}

\author{Wenhua Ling\orcidlink{0009-0009-9142-6608}}
\affiliation{Australia Telescope National Facility, CSIRO, Space \& Astronomy, PO Box 76, Epping, 1710, NSW, Australia}

\author{Rami Mandow\orcidlink{0000-0001-5131-522X}}
\affiliation{Australia Telescope National Facility, CSIRO, Space \& Astronomy, PO Box 76, Epping, 1710, NSW, Australia}
\affiliation{Department of Mathematics and Physical Sciences, Macquarie University, NSW 2109, Australia}

\author{Saurav Mishra\orcidlink{0009-0001-5633-3512}}
\affiliation{ARC Centre for Excellence for Gravitational Wave Discovery (OzGrav), Hawthorn, VIC, 3122, Australia} 
\affiliation{Australia Telescope National Facility, CSIRO, Space \& Astronomy, PO Box 76, Epping, 1710, NSW, Australia}
\affiliation{Centre for Astrophysics and Supercomputing, Swinburne University of Technology, P.O. Box 218, Hawthorn, VIC 3122, Australia}

\author{Daniel J. Reardon\orcidlink{0000-0002-2035-4688}}
\affiliation{Centre for Astrophysics and Supercomputing, Swinburne University of Technology, P.O. Box 218, Hawthorn, VIC 3122, Australia}

\author{Christopher J Russell\orcidlink{0000-0002-1942-7296}}
\affiliation{CSIRO Scientific Computing, Australian Technology Park, Locked Bag 9013, Alexandria, NSW 1435, Australia}

\author{Ryan M. Shannon\orcidlink{0000-0002-7285-6348}}
\affiliation{Centre for Astrophysics and Supercomputing, Swinburne University of Technology, P.O. Box 218, Hawthorn, VIC 3122, Australia}

\author{Jacob Cardinal Tremblay\orcidlink{0000-0001-9852-6825}}
\affiliation{Max Planck Institute for Gravitational Physics (Albert Einstein Institute), 30167 Hannover, Germany}
\affiliation{Leibniz Universität Hannover, 30167 Hannover, Germany}

\author{Jingbo Wang\orcidlink{0000-0001-9782-1603}}
\affiliation{Institute of Optoelectronic Technology, Lishui University, Lishui 323000, China}

\author{Lei Zhang\orcidlink{0000-0001-8539-4237}}
\affiliation{Centre for Astrophysics and Supercomputing, Swinburne University of Technology, P.O. Box 218, Hawthorn, VIC 3122, Australia}
\affiliation{National Astronomical Observatories, Chinese Academy of Sciences, A20 Datun Road, Chaoyang District, Beijing 100101, China}

\author{Andrew Zic\orcidlink{0000-0002-9583-2947}}
\affiliation{ARC Centre for Excellence for Gravitational Wave Discovery (OzGrav), Hawthorn, VIC, 3122, Australia} 
\affiliation{Australia Telescope National Facility, CSIRO, Space \& Astronomy, PO Box 76, Epping, 1710, NSW, Australia}


\begin{abstract}
In recent years, several pulsar timing array collaborations have reported evidence for a nanohertz gravitational wave background (GWB). Such a background signal could be produced by supermassive binary black holes, early-Universe processes such as inflation and phase transitions, or a mixture of both. One way to disentangle different contributions to the GWB is to search for anisotropic signatures. In this work, we search for anisotropy in the GWB using the third data release of the Parkes Pulsar Timing Array. Our analysis employs both the radiometer method and the spherical harmonic basis to characterize the distribution of GWB power across the sky. We calculate the angular power in the lowest five frequency bins and compare it with detection thresholds determined under the null hypothesis of isotropy. In the 5.26 nHz frequency bin, we identify a hotspot in the reconstructed sky map with a $p$-value of $0.016$ (the lowest in our analysis), which we attribute to noise fluctuations.
While our search reveals no statistically significant anisotropy, we expect that the precise measurement of angular power spectrum of the GWB will become instrumental in determining the origin of the nanohertz GWB signal. 

\end{abstract}

\maketitle

\section{Introduction}

Complimentary to ground-based interferometers such as LIGO \cite{LIGOScientific:2014pky}, Virgo \cite{VIRGO:2014yos}, and KAGRA \cite{KAGRA:2018plz}, pulsar timing arrays (PTAs) have emerged as a powerful tool to detect gravitational waves (GWs) in the nanohertz frequency band.
Currently, several regional PTAs have accumulated around two decades of high-precision timing observations of millisecond
pulsars (MSPs), including the Parkes Pulsar Timing Array (PPTA;~\citep{Manchester_2013}), the North American Nanohertz Observatory for Gravitational Waves (NANOGrav;~\citep{Ransom_2019}) and the European Pulsar Timing Array (EPTA;~\citep{Desvignes_2016}). Together with the Indian Pulsar Timing Array (InPTA;~\citep{Joshi_2018}), they form the International Pulsar Timing Array (IPTA;~\citep{Verbiest_2016}). In addition, the Chinese Pulsar Timing Array (CPTA;~\citep{Lee_2016}) and the MeerKAT Pulsar Timing Array (MPTA;~\citep{Miles_2022}) have been conducting independent observations for $\gtrsim \unit[5]{years}$, and data-sharing agreements between MPTA/CPTA and the IPTA are also in place. Recently, these PTAs have all reported evidence supporting the existence of a GW background (GWB)~\citep{Agazie_2023a,Reardon_2023,EPTA_InPTA_2023,Xu_2023,Miles_2024}.

In the PTA frequency band, the GWB may arise from a variety of potential sources, with either astrophysical or cosmological origin. The dominant astrophysical contribution is expected to come from a cosmic population of supermassive binary black holes (SMBBHs) with masses approximately $10^{8}-10^{10} M_{\odot} $, during their slow inspiral phase~\citep{Begelman_1980,Jaffe_2003,Sesana_2009}. On the other hand, cosmological GWB sources are typically associated with processes in the early Universe, and their underlying mechanisms are more diverse~\citep{Caprini_2018}. These include inflation~\citep{Vagnozzi_2021,Niu_2023}, cosmological phase transitions~\citep{Caprini_2010,Ellis_2020,Xue_2021}, cosmic strings~\citep{Ellis_2021,Chen_2022,Bian_2022}, domain walls~\citep{Kawasaki_2011,Ferreira_2023}, and scalar-induced GWs~\citep{Domenech_2021}. Determining the origin of the GWB is a primary scientific goal of PTA experiments in the near future.

Most current PTA analyses have targeted an isotropic GWB. However, a GWB generated by a finite number of inspiraling SMBBHs is expected to exhibit some degree of anisotropy, arising from their spatial clustering and uneven distribution over cosmological scales.
Certain cosmological mechanisms can also produce anisotropic GWBs, although their resulting anisotropy signatures are somewhat different. For example, superhorizon-scale inhomogeneities~\citep{Liu:2020mru}, preheating dynamics~\citep{Bethke:2013aba,Bethke:2013vca}, or non-Gaussian curvature perturbations~\citep{Bartolo:2019zvb,Yu:2023jrs} can imprint significant anisotropies on the GWB. Consequently, the detection of anisotropy cannot be attributed solely to an astrophysical origin. The precise angular power spectrum of the GWB will therefore serve as a critical discriminant between the SMBBH population model and exotic cosmological mechanisms.

A number of studies have been devoted to exploring the capabilities of PTAs to detect GWB anisotropy. Refs. \cite{Mingarelli_2013,Gair_2014,Hotinli_2019} calculated the response of pulsar timing residuals to anisotropic GWB and parameterized the signal angular distribution through spherical harmonic expansion. This method can effectively extract the directional dependency information of GWBs and provide key clues for understanding their sources. In addition, Refs. \cite{Taylor_2013,Taylor_2020} developed a set of Bayesian analysis techniques and verified them using simulated GWBs.

To date, three major PTA collaborations have carried out searches for anisotropy in the GWB. The earliest such effort was by the EPTA collaboration~\cite{Taylor_2015}. Using timing data from the six most precisely timed pulsars, they placed constraints on the anisotropy and found that in a background distribution with multipoles $ l>0 $, the $ 95\% $ upper limit on the multipolar strain amplitude is $ \lesssim 40\% $ of the monopole strain amplitude. More recently,~\citet{Agazie_2023c} employed both Bayesian and frequentist methods, using different basis functions to characterize the angular power and search for anisotropic features in the GWB.
\citet{Grunthal_2024} used spherical harmonic decomposition to search for anisotropy in the MPTA data and found a potential hotspot in their $7$ nHz clean map, with a corresponding $p$-value of $0.015$. So far, no conclusive evidence for anisotropy in the GWB has been found in PTA data sets.

In this paper, we adopt a frequentist framework to search for anisotropy in the GWB using the PPTA Data Release 3 (DR3). The remainder of this paper is organized as follows. In \Sec{method}, we describe the theoretical framework for reconstructing the GWB power distribution using different basis functions, and outline the detection statistics employed to evaluate the statistical significance of any anisotropy. In \Sec{results}, we present the main analysis results based on PPTA DR3, including measurements of the angular power spectrum in the lowest five frequency bins, reconstructions of the GWB sky map, and $p$-value calculations. Finally, in \Sec{conclusion}, we summarize and interpret our results, and discuss future prospects of anisotropy search using PTAs.

\section{Methods} \label{method}
To search for anisotropic signatures of the GWB, it is necessary to select appropriate basis functions on the celestial sphere to expand and represent the GWB power distribution. Different basis choices offer sensitivity to different angular scales. In this work, we adopt both the radiometer pixel basis \citep{Ballmer_2006,Mitra_2008,Cornish_2014} and the spherical harmonic basis \citep{Mingarelli_2013,Taylor_2013,Gair_2014,Taylor_2015} to measure the angular power distribution of the GWB.

\subsection{Characterization of the GWB anisotropy}

\subsubsection{Radiometer pixel basis}
When aiming to search for a localized bright point source on the sky, the radiometer pixel basis provides an appropriate framework.
It allows an independent reconstruction of the GWB power in each equal-area pixel, which can be written as
\begin{equation}
    P(\hat{\Omega}) = \sum_{\hat{\Omega}'} P_{\hat{\Omega}'} \delta^2(\hat{\Omega}, \hat{\Omega}'),
\end{equation}
where $\hat{\Omega}$ denotes a sky direction and $P_{\hat{\Omega}'}$ is the power of the signal in each pixel. The number of pixels in the sky map is determined by $N_{\text{pix}} = 12 N_{\text{side}}^2$, where $N_{\text{side}}$ specifies the resolution of the HEALPix grid~\citep{Górski_2005}. The achievable angular resolution is constrained by the number of pulsar pairs $N_{\text{cc}}$ in the PTA dataset. Following the empirical criterion $N_{\text{pix}} \leq N_{\text{cc}}$~\citep{Cornish_2014,Romano_2017}, and considering the 30 MSPs selected from the PPTA DR3 dataset, we obtain $N_{\text{cc}} = 435$, corresponding to $N_{\text{side}} \leq 4$.

\subsubsection{Spherical harmonic and Square-root spherical harmonic basis}

For a diffuse GWB, such as one dominated by dipolar or quadrupolar distribution, a decomposition in the spherical harmonic basis is appropriate. In this case, the GW power distribution across the sky can be expanded in terms of spherical harmonics~\citep{Allen_1997,Cornish_2001,Kudoh_2005,Thrane_2009}
\begin{equation}
    P(\hat{\Omega}) = \sum_{l=0}^{\infty} \sum_{m=-l}^{l} c_{lm} Y_{lm}(\hat{\Omega}),
\end{equation}
where $Y_{lm}$ are the spherical harmonic functions that encode the angular structure on the sky, and $c_{lm}$ are the mode coefficients that quantify the contribution of each multipole moment to the GWB power distribution.

While decomposing the GWB power directly into spherical harmonics is a natural choice, it allows the power to take negative values, which is clearly unphysical. To address this issue, an alternative approach involves performing a spherical harmonic expansion of the square root of the GWB power. This square-root parameterization ensures by construction that the reconstructed power remains positive across the sky. This technique was originally developed within a Bayesian framework: for instance, \citet{Payne_2020} implemented it for LIGO, and \citet{Banagiri_2021} also extended it to LISA. \citet{Taylor_2020} introduced the square-root expansion into PTA analyses using Bayesian techniques to reconstruct the GWB angular power distribution. Building on this, \citet{Pol_2022} applied the method within a frequentist framework for PTA data. More recently, \citet{Agazie_2023c} incorporated this technique into PTA analyses, employing both frequentist and Bayesian approaches. This parameterization can be expressed as follows
\begin{equation}
    P(\hat{\Omega}) = \left[P(\hat{\Omega})^{1/2}\right]^2 = \left[ \sum_{L=0}^{\infty} \sum_{M=-L}^{L} a_{LM} Y_{LM}(\hat{\Omega}) \right]^2,
\end{equation}
where $Y_{LM}$ are the real-valued spherical harmonics, and $a_{LM}$ denotes the search coefficients. We have capitalized the component labels ($LM$) to be distinct from the usual ($lm$) components of the GWB power. The coefficients of the linear spherical harmonic decomposition are related to the search coefficients through
\begin{equation}
    c_{lm} = \sum_{LM} \sum_{L'M'} a_{LM} a_{L'M'} \beta^{LM,L'M'}_{lm},
\end{equation}
where $\beta_{lm}^{LM,L'M'}$ is defined as
\begin{equation}
    \beta_{lm}^{LM,L'M'} = \sqrt{\frac{(2L + 1)(2L' + 1)}{4\pi (2l + 1)}} C_{LM,L'M'}^{lm} C_{L0,L'0}^{l0} ,
\end{equation}
with $C_{LM,L'M'}^{lm}$ representing the Clebsch–Gordon coefficients. This construction ensures that the reconstructed GWB power distribution remains positive by effectively constraining the allowed forms of $c_{lm}$.  

The distribution of anisotropy across different angular scales is quantified by the power spectrum coefficients
\begin{equation}
    C_l = \frac{1}{2l + 1} \sum_{m=-l}^{l} |c_{lm}|^2.
\end{equation}
Thus, the angular power spectrum $C_l$ quantifies the statistical fluctuations of the GWB power on angular scales corresponding to $\theta = 180^\circ / l$. In the case of an isotropic GWB, all the power is concentrated in the monopole component ($l = 0$), indicating a uniform distribution of GWB power across the sky. Conversely, the presence of anisotropy leads to power appearing at higher multipoles ($l > 0$). As $l$ increases, the power distribution of the GWB will extend to smaller spatial scales.

As discussed by~\citet{Boyle_2012}, the maximum angular resolution achievable in a search for anisotropy is determined by the diffraction limit, setting the highest multipole $l_{\max}$ that a PTA can probe. This limit scales as $l_{\max} \sim \sqrt{N_{\mathrm{psr}}}$~\citep{Romano_2017}. For the PPTA DR3 analyzed here, this leads to $l_{\max} = 5$, corresponding to an angular resolution of $\theta = 36^\circ$. In contrast, the radiometer basis adopted in our analysis achieves an angular resolution of $14.6^\circ$. Thus, the spherical harmonic and radiometer bases are sensitive to anisotropies at different angular scales. By combining these two approaches, we can more comprehensively characterize potential anisotropic features in the GWB.

\subsection{Maximum likelihood estimates of the GWB power}

In PTA data analysis, we work with the timing residuals $\delta \boldsymbol{t}$, produced by subtracting TOAs predicted by a timing model $\boldsymbol{t}_{\text{det}}$ from the measured TOAs $\boldsymbol{t}_{\text{TOA}}$, i.e., $\delta \boldsymbol{t} = \boldsymbol{t}_{\text{TOA}} - \boldsymbol{t}_{\text{det}}$. By taking the expectation values of the product of different residuals, we construct covariance matrices
\begin{equation}
   \boldsymbol{C}_a = \langle \delta \boldsymbol{t}_a \delta \boldsymbol{t}_a^T \rangle,
\end{equation}
\begin{equation}
   \boldsymbol{S}_{ab} = \langle \delta \boldsymbol{t}_a \delta \boldsymbol{t}_b^T \rangle \Big|_{a \neq b},
\end{equation}
where $\boldsymbol{C}_a$ is the measured autocovariance matrix, and $\boldsymbol{S}_{ab}$ is the cross-covariance matrix. Here, the vectors $\delta \boldsymbol{t}_a$ and $\delta \boldsymbol{t}_b$ correspond to the timing residuals of pulsars $a$ and $b$, respectively.

The cross-correlation coeﬃcients $\rho_{ab}$ for each pulsar pair can be expressed as~\citep{Demorest_2013,Siemens_2013,Chamberlin_2015,Vigeland_2018,Pol_2022}
\begin{equation}
    \rho_{ab} = \frac{\delta \boldsymbol{t}_a^T \boldsymbol{C}_a^{-1} \hat{\boldsymbol{S}}_{ab} \boldsymbol{C}_b^{-1} \delta \boldsymbol{t}_b^T}{\text{tr}[\boldsymbol{C}_a^{-1} \hat{\boldsymbol{S}}_{ab} \boldsymbol{C}_b^{-1} \hat{\boldsymbol{S}}_{ba}]},
\end{equation}
where $\hat{\boldsymbol{S}}_{ab}$ is defined so that $A_{\text{gw}}^2 \Gamma_{ab} \hat{\boldsymbol{S}}_{ab} = \boldsymbol{S}_{ab}$. 
Here $A_{\rm gw}$ is the GWB amplitude parameter of for the assumed strain spectrum template,
and $\Gamma_{ab}$ is the expected GWB-induced cross-correlation between pulsars $a$ and $b$
(e.g., the Hellings-Downs (HD) factor for an isotropic GWB).

The uncertainty $\sigma_{ab}$ on the correlation coeﬃcients is given by
\begin{equation}
    \sigma_{ab} = \left( \text{tr}[\boldsymbol{C}_a^{-1} \hat{\boldsymbol{S}}_{ab} \boldsymbol{C}_b^{-1} \hat{\boldsymbol{S}}_{ba}] \right)^{-1/2},
\end{equation}

The measured cross-correlations between pulsar pairs in a PTA can be modeled using the so-called overlap reduction function (ORF)~\citep{Gair_2014,Taylor_2020}, which can be recast into a general matrix form as 
\begin{equation}
    \boldsymbol{\Gamma} = \boldsymbol{R} \boldsymbol{P},
\end{equation}
where $\boldsymbol{\Gamma}$ is a vector containing the ORF values for each distinct pulsar pair, $\boldsymbol{P}$ is the vector describing the GWB power distribution across the sky, and the overlap response matrix $\boldsymbol{R}$ is given by
\begin{equation}
R_{ab,k} = \frac{3}{2N_{\text{pix}}} \left[ \mathcal{F}^+_{a,k} \mathcal{F}^+_{b,k} + \mathcal{F}^\times_{a,k} \mathcal{F}^\times_{b,k} \right],
\end{equation}
where $N_{\mathrm{pix}}$ denotes the total number of sky pixels in the map. Here, $\mathcal{F}^A_{a,k}$ denotes the antenna response function of pulsar $a$ to a GW with polarization $A$ ($A \in \{+, \times\}$) arriving from the sky direction corresponding to pixel $k$,
the detailed form of which can be found in \citep{Taylor_2020,Pol_2022}.

Assuming that the noise in the cross-correlation measurements is stationary Gaussian, the estimator can be constructed by maximizing the following likelihood function~\citep{Mitra_2008,Thrane_2009,Pol_2022}
\begin{equation}
    p(\boldsymbol{\rho} | \boldsymbol{P}) = \frac{\exp \left[ -\frac{1}{2} (\boldsymbol{\rho} - \boldsymbol{R} \boldsymbol{P})^{T} \boldsymbol{\Sigma}^{-1} (\boldsymbol{\rho} - \boldsymbol{R} \boldsymbol{P}) \right]}{\sqrt{\det(2\pi \boldsymbol{\Sigma})}},
\end{equation}
where $\boldsymbol{\Sigma}$ is the diagonal noise covariance matrix of the cross-correlation measurements, with shape $N_\text{cc} \times N_\text{cc}$. Here $N_{\text{cc}} = N_{\text{psr}} (N_{\text{psr}} - 1) / 2$, where $N_{\text{psr}}$ is the number of pulsars in the array.

The maximum likelihood estimator for the GWB power distribution is given by~\citep{Thrane_2009, Romano_2017, Ivezić_2019, Pol_2022}
\begin{equation}
\hat{\boldsymbol{P}} = \boldsymbol{M}^{-1} \boldsymbol{X},
\end{equation}
where 
\begin{equation}
\boldsymbol{M} = \boldsymbol{R}^{T} \boldsymbol{\Sigma}^{-1} \boldsymbol{R},
\end{equation}
is the Fisher information matrix and
\begin{equation}
\boldsymbol{X} = \boldsymbol{R}^{T} \boldsymbol{\Sigma}^{-1} \boldsymbol{\rho},
\end{equation}
is referred to as the dirty map, an inverse noise-weighted estimate of the GW power distribution on the sky as seen through the response of the pulsars~\citep{Thrane_2009,Ali-Haimoud_2021,Pol_2022,Agazie_2023c}.

Note that in the square-root spherical harmonic basis, the maximum likelihood solution cannot be found analytically. Instead, it can be computed numerically~\citep{Levenberg_1994,Marquardt_1963,Newville_2021}, as implemented in the \texttt{MAPS} package~\citep{Pol_2022}.

When we are interested in point sources or localized features in the sky, instead of trying to reconstruct the entire GW sky map, we can focus on analyzing discrete pixels in the sky, each pixel representing the signal strength in a specific direction. This approach is well suited for non-uniform signal distributions, such as the detection of local hotspots. Although the radiometer map~\citep{Mitra_2008,Thrane_2009} is less sensitive to large-scale anisotropies, it is suitable for visualizing local hotspots or small-scale structures. The radiometer map is given by
\begin{equation}
\boldsymbol{P}_{\hat{\Omega}} = \boldsymbol{M}_{\hat{\Omega} \hat{\Omega}}^{-1} \boldsymbol{X}_{\hat{\Omega}} ,
\end{equation}
with an associated uncertainty given by
\begin{equation}
\boldsymbol{\sigma}^{P}_{\hat{\Omega}} = \left( \boldsymbol{M}_{\hat{\Omega} \hat{\Omega}} \right)^{-1/2}.
\end{equation}
where the subscript $\hat{\Omega}$ indicates that it works on a pixel basis. In the radiometer basis, where each pixel is modeled independently, the power in each pixel can be obtained by taking the inverse of the corresponding diagonal element of the Fisher matrix, rather than inverting the full matrix.

\subsection{Statistical significance} \label{statistics}
In the final step of anisotropy search, we aim to determine whether there is statistically significant evidence of anisotropy. This process involves defining a suitable detection statistic and comparing it against the distribution expected under the null hypothesis, which states that the GWB is isotropic. Consequently, any statistically significant deviation from the null hypothesis would indicate potential anisotropic features in the data.


For the square-root spherical harmonics basis, a commonly used test statistic is the signal-to-noise ratio (SNR) defined by the maximum likelihood ratio~\citep{Pol_2022,Agazie_2023c}
\begin{equation}
\mathrm{S/N} = \sqrt{2 \ln \left[ \frac{p(\boldsymbol{\rho} | \boldsymbol{P})}{{p(\boldsymbol{\rho} | \boldsymbol{P}_{\text{iso}})}}\right]},
\label{eq:SNR}
\end{equation}
where $\boldsymbol{P}$ denotes the reconstructed GWB power distribution under the anisotropic model, while $\boldsymbol{P}_{\text{iso}}$ corresponds to a constant power distribution across the sky, representing an isotropic background with $\boldsymbol{P}_{\text{iso}} = 1$. The statistical significance of the measured SNR is evaluated by comparing it against a null-hypothesis distribution. This null distribution is constructed by generating multiple realizations of an isotropic GWB. For each pulsar pair, mock cross-correlations are sampled from Gaussian distributions centered on the HD expectation, with standard deviations determined by the measured cross-correlation uncertainties. These mock datasets are then processed through the same frequentist pipeline to compute the associated SNRs. The ensemble of SNRs obtained in this way forms the null distribution, against which the observed SNR is compared to derive the corresponding $p$-value. A $p$-value smaller than $3\times10^{-3}$ (corresponding to approximately $3\sigma$ significance) would suggest statistically significant evidence of anisotropy.

For the radiometer pixel basis, the SNR in each pixel is defined as
\begin{equation}
\mathrm{S/N} = \frac{\boldsymbol{P}_{\hat{\Omega}}}{\boldsymbol{\sigma}^{P}_{\hat{\Omega}}}.
\end{equation}
We compute the maximum and minimum SNR across all pixels in the reconstructed sky map for each realization under the null hypothesis, thereby constructing the null distributions of these statistics. We then compare the observed maximum and minimum SNR values with the corresponding null distributions to obtain their associated $p$-values. A $p$-value below $3 \times 10^{-3}$ would indicate that the signal deviates significantly from isotropy, thus providing evidence for anisotropy in the GWB.


In addition to the calculation of SNRs, we also compute the detection threshold for the square-root spherical harmonics basis.
Under the null hypothesis that the GWB is isotropic, a threshold angular power value $C_l^{\mathrm{th}}$ is determined for each multipole, corresponding to a $p$-value of $3 \times 10^{-3}$. If the measured angular power at any multipole exceeds this threshold, it is considered to be a statistically significant detection of anisotropy at approximately the $3\sigma$ level. This method allows for a detailed assessment of anisotropy at individual angular scales, while the SNR statistic offers a global assessment of anisotropy.

\begin{table*}[tbp]  
    \centering
    \caption{\label{tab1} 
    The $p$-values for the lowest five frequency bins, computed using both the square-root spherical harmonic basis and the radiometer pixel basis. The $p$-values are obtained by comparing the measured SNR distributions with those expected under the null hypothesis of an isotropic GWB.
    }
    \begin{ruledtabular}  
    \begin{tabular}{c c c c}
        Frequency (nHz) & \multicolumn{1}{c}{Square-root spherical harmonic} & \multicolumn{2}{c}{Radiometer} \\ 
        \cline{2-2} \cline{3-4}  
        & $p$ & $p({\rm SNR_{max})}$ & $p({\rm SNR_{min}})$ \\ 
        \hline
        1.75 & 0.50  & 0.56 & 0.90 \\
        3.50 & 0.72  & 0.66 & 0.70 \\
        5.26 & 0.021 & 0.016 & 0.11 \\
        7.01 & 0.57  & 0.44 & 0.83 \\
        8.76 & 0.51  & 0.12 & 0.65 \\
    \end{tabular}
    \end{ruledtabular}
\end{table*}


\begin{figure}
    \begin{center}
        \includegraphics[width=0.95\columnwidth]{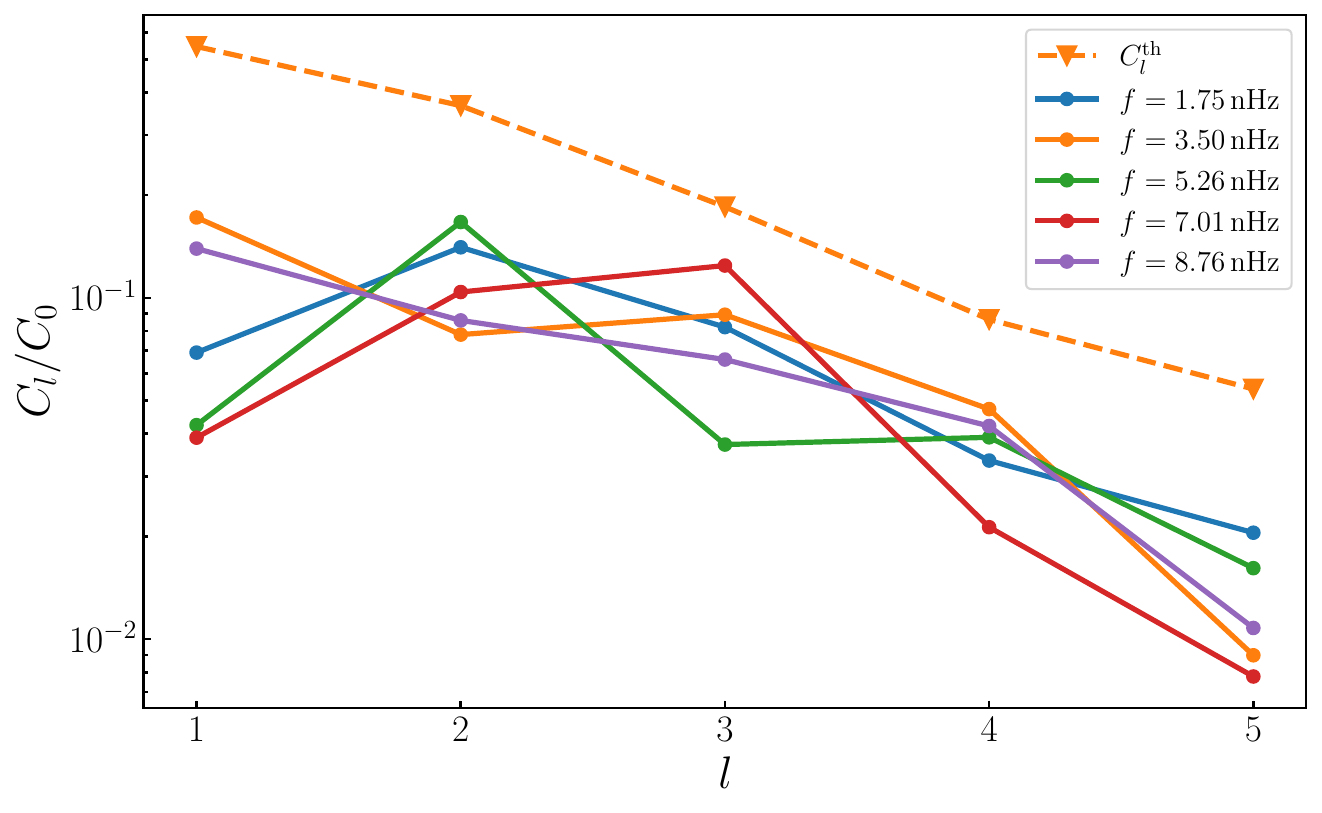}
        \caption{The angular power spectra measured using the square-root spherical harmonics basis.   The orange dashed line indicates the detection threshold, corresponding to a significance level of $p = 3\times 10^{-3}$ for rejecting the isotropy hypothesis. The colored solid line shows the measured angular power spectrum in the lowest five frequency bins.
       } 
        \label{Cl_per}  
    \end{center}
\end{figure}

\section{Results} \label{results}
The PPTA DR3 contains observations of 32 MSPs taken with the 64-meter Parkes radio telescope (Murriyang), spanning up to 18 years with a typical observation cadence of approximately three weeks.
Following \citet{Reardon_2023} and \citet{ZhaoSY25ppta-cw}, we exclude PSR
J1824$-$2452A (due to strong intrinsic red noise) and PSR J1741+1351 (due to a small number of observations) in this work.
More details about the PPTA DR3 and the noise properties can be found in Refs. \citep{Zic_2023,reardon2023noises}.
We present in \Fig{sensitivity} of Appendix \ref{sensitivity_map} a sensitivity map of the PPTA.


\subsection{Square-root spherical harmonic basis}

Using the square-root spherical harmonic basis, we employ the recently developed per-frequency optimal statistic \citep{Gersbach_2025_PFOS} to compute the angular power of the GWB in the lowest five frequency bins, and compare the results with the corresponding detection thresholds to assess the significance of any anisotropic signal, as shown in \Fig{Cl_per}. The per-frequency analysis indicates that, across the range of multipole moments considered, the measured angular power in each frequency bin remains below the detection thresholds, showing no statistically significant evidence of frequency-dependent anisotropy. Some variations are observed in the independent estimates across different frequency bins, which may be influenced by specific frequency-domain features or noise fluctuations.

In the lowest five frequency bins, we compare the measured anisotropic SNRs with the distribution under the null hypothesis and obtain the corresponding p-values, which are shown in the “Square-root spherical harmonic” column of \Table{tab1}. It is worth noting that in the 5.26 nHz frequency bin, the p-value of $p=0.021$ is significantly smaller than other frequency bins. 
We note that the $p$-values presented in \Table{tab1} represent the global significance of the $l=1-5$ anisotropic model. The relatively low $p$-value at 5.26 nHz is driven by the presence of a localized hotspot—which is also identified as the most significant feature in our radiometer search ($p=0.016$). In contrast, an isolated elevation in a single multipole, such as the $l=3$ component at 7.01 nHz (as shown in \Fig{Cl_per}), does not result in a similarly small global $p$-value because it lacks a corresponding localized structure.


We also measure the angular power of the GWB based on the power-law spectrum template (for the entire band) and calculate the corresponding $p$-value. These results are provided in Appendix~\ref{power-law}.

\subsection{Radiometer pixel basis}

\begin{figure*}[htbp]
    \begin{center}
        \includegraphics[width=\linewidth]{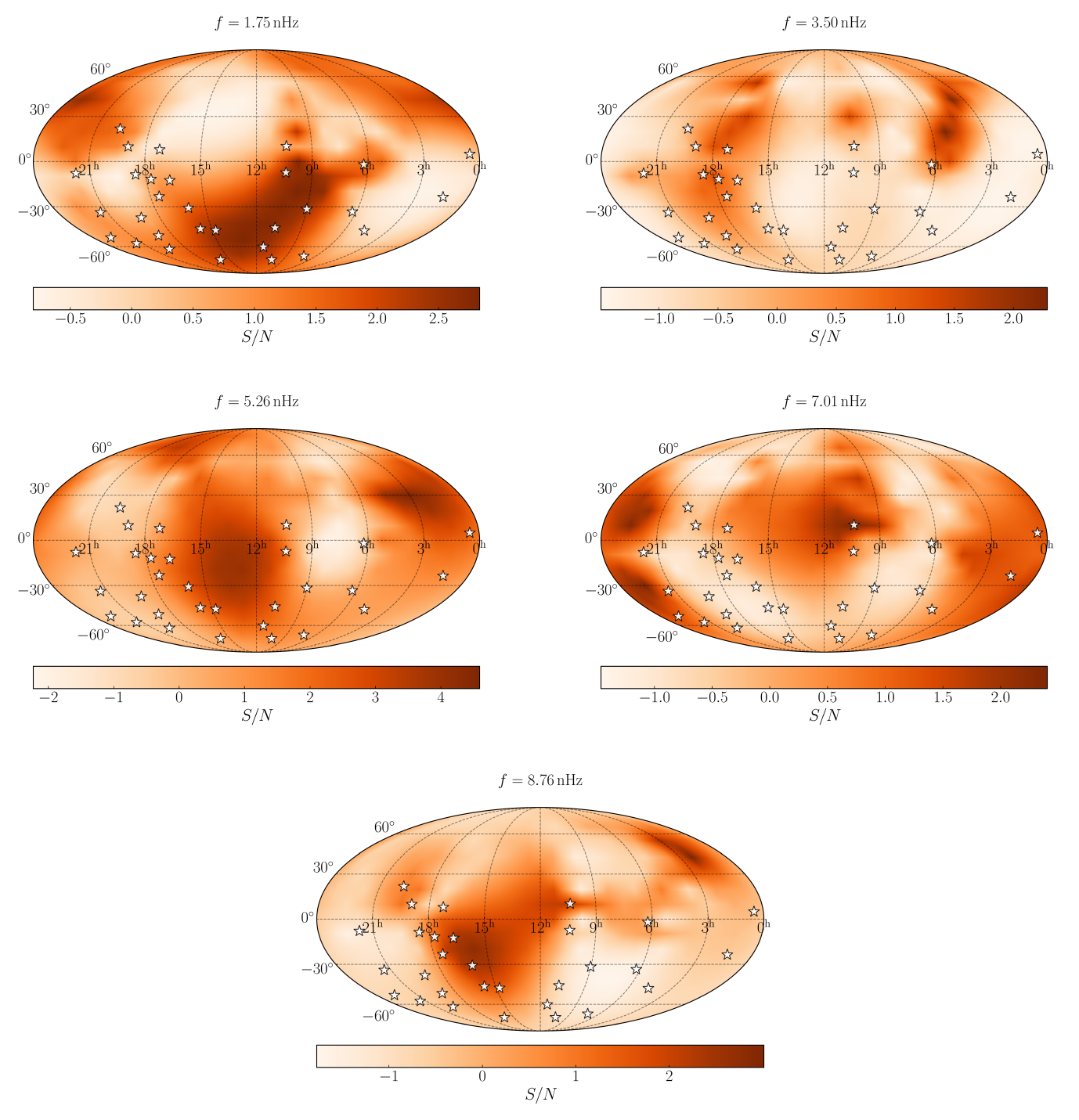}
        \caption{Sky maps of the anisotropy SNR of the GWB reconstructed in the lowest five frequency bins using the radiometer method. Each panel corresponds to a different frequency bin, and the white stars indicate the positions of PPTA pulsars. A potential hotspot appears in the 5.26 nHz frequency bin, with the maximum SNR corresponding to a $p$-value of 0.016.
        } 
        \label{radio_sn_five}  
    \end{center}
\end{figure*}

\Fig{radio_sn_five} presents the GWB sky maps reconstructed in the lowest five frequency bins using the radiometer method. We calculate the $p$-values corresponding to the maximum and minimum SNR values in each map, with the results listed in the “Radiometer” column of \Table{tab1}. The $p$-values for the minimum SNR range from 0.11 to 0.90, with no abnormally low values observed, indicating the robustness and reliability of our analysis. Most maximum-SNR $p$-values are close to 1, suggesting no statistically significant deviation from isotropy across the frequency bands analyzed. It is also worth noting that in the 5.26 nHz band, the $p$-value corresponding to the maximum SNR is 0.016, which is consistent with results from our spherical harmonic analysis.

This finding highlights a key consideration for anisotropic searches: in realistic scenarios, a localized contribution (e.g., from a nearby binary) may be superimposed on an unresolved, approximately isotropic stochastic GWB. In our framework, this corresponds to a sky power distribution consisting of a dominant isotropic component plus a localized excess. Our pixel-basis analysis remains applicable in this context as a search for localized departures from such a background.

However, an unresolved background induces common residuals across pulsars that are correlated according to the HD curve. In the present frequentist implementation, we approximate the covariance of the pairwise cross-correlation measurements as diagonal. As discussed in recent methodological advancements (e.g., \cite{Gersbach_2025}), a more complete treatment—incorporating the full inter-pair covariance and noise marginalization—would properly account for the isotropic background's correlated noise across all pulsar pairs and could refine the resulting pixel S/N and associated significance. We leave the implementation of such a next-generation anisotropic optimal statistic for future work.

To further investigate the 5.26 nHz feature, we compare its sky location and frequency with results from continuous-wave searches that targeted individual SMBBH sources.
Also based on PPTA DR3, \citet{ZhaoSY25ppta-cw} reported that at 5.26 nHz the continuous wave signal hypothesis was disfavored in both Bayesian and frequentist analysis.
Furthermore, we
cross-check the hot spot candidate against the 114 SMBBH candidates reported by \citet{ng_target} using NANOGrav data, and find that none of these candidates overlaps with the hotspot in either GW frequency or sky location.

\section{Summary and Conclusions} \label{conclusion}

We search for anisotropy in the GWB using PPTA DR3. Adopting a frequentist analysis method, we parameterize the sky map of the GWB power using both the square-root spherical harmonic and radiometer pixel bases. 
We first perform a per-frequency analysis of the lowest five frequency bins, allowing independent estimation of the GWB power in different frequencies.
Across these frequencies, our results show no statistically significant evidence for anisotropy. The angular power spectra measured under the square-root spherical harmonic parameterization remain consistently below the detection threshold.
We also perform an analysis assuming a power-law spectrum for the GWB and find no significant evidence of anisotropy.

A noteworthy finding from this work is a potential hotspot at 5.26~nHz, with a $p$-value of $0.016$, identified in the radiometer search\footnote{This $p$-value is similar to the potential hotspot (with $p=0.015$) found in the 7 nHz sky map of MPTA data \citep{Grunthal_2024}.}.
This is the lowest $p$-value found across our analyses; a comparable $p$-value of $0.021$ is found in the spherical harmonic basis analysis.
We think it is likely due to noise fluctuations because 1) the location of the hotspot is in the least sensitive sky region of PPTA DR3 (see Fig. \ref{sensitivity}), 2) no support for continuous GWs from previous searches for individual SMBBHs was found at its frequency and sky location, and 3) the $p$-value is below the threshold of $ 3 \times 10^{-3}$.
With the improved sensitivity and sky coverage from future datasets such as IPTA DR3, it will be interesting to revisit these features.

Looking into the future, the imminent third data release from the IPTA and the advent of the Square Kilometre Array (SKA) will provide the critical tools to dissect the nanohertz GWB signal and map its structure across the sky. These advancements will deliver a transformative trifecta of capabilities: longer timing baselines that refine our spectral sensitivity, an expanded and spatially diverse pulsar array that acts as a high-resolution interferometer, and more homogeneous sky coverage that is essential for clean angular decomposition. This collective leap in precision will enable us to move beyond a mere detection of the GWB's amplitude. We will, for the first time, reconstruct detailed sky maps of the nanohertz GW sky.
The specific pattern of these anisotropies is highly informative.
Large-scale patterns could trace the uneven distribution of SMBBHs across the cosmic web of galaxies or even reveal our Solar System's motion through the background.
Small-scale, bright spots would strongly suggest the presence of a single, nearby, or exceptionally loud SMBBH.
If multiple sources (both astrophysical and cosmological) contribute to the GWB, characterizing their distinct anisotropic patterns and spectral signatures offers the most promising method to disentangle them.
Therefore, mapping the GWB's anisotropy is not a secondary task; it is the fundamental key to determining its source, distinguishing a cosmic hum of black holes from more exotic physics beyond the Standard Model.
\begin{acknowledgments}
This work is supported by the National Key Research and Development Program of China (No. 2023YFC2206704), the National Natural Science Foundation of China (Grant No.~12203004), the Fundamental Research Funds for the Central Universities, and the Supplemental Funds for Major Scientific Research Projects of Beijing Normal University (Zhuhai) under Project ZHPT2025001.
The Parkes radio telescope (Murriyang) is part of the Australia Telescope National Facility which is funded by the Australian Government for operation as a National Facility managed by CSIRO. We acknowledge the Wiradjuri People as the traditional owners of the Observatory site.
Parts of this work were funded through the ARC Centre of Excellence for Gravitational Wave Discovery (CE230100016). 
This work is partly supported by the National Natural Science Foundation of China under Grant No.~12405056, the Natural Science Foundation of Hunan Province under Grant No.~2025JJ40006, and the Innovative Research Group of Hunan Province under Grant No.~2024JJ1006.

\end{acknowledgments}

\emph{Facilities:} Parkes

\emph{Software:} \texttt{enterprise} \citep{enterprise}, \texttt{enterprise\_extensions} \citep{enterprise_ext}, 
\texttt{lmfit} \citep{Newville_2021}, 
\texttt{healpy} \citep{Zonca_2019_healpy}, 
\texttt{HEALPix} \citep{Górski_2005}, 
\texttt{Defiant} \citep{Gersbach_2025_PFOS}, 
\texttt{MAPS} \citep{Pol_2022}.




\appendix
\setcounter{figure}{0}
\renewcommand{\thefigure}{S\arabic{figure}}

\section{The sensitivity map}\label{sensitivity_map}

In order to quantify the detection capability of PPTA DR3 for GWB signals in different sky directions, we plot a sensitivity sky map based on the radiometer method. The definition of sensitivity is derived from the Fisher information matrix in each sky direction, and its diagonal elements reflect the ability of the PTA to constrain the signal power in each pixel direction. Specifically, a larger diagonal element of the Fisher matrix indicates smaller uncertainty in that direction, and thus higher sensitivity.

The response of each pulsar pair $(a, b)$ is described by the sky integral in the ORF, which accounts for the combined response of the pair to GWs incident from all sky directions. Due to differences in observational baselines, noise characteristics, and sky locations among the pulsars in the PTA, the information contributed by each pulsar pair varies. The Fisher information matrix aggregates these contributions in the pixel basis, such that the diagonal elements can be used to evaluate the precision of GWB power estimates in specific sky directions.

\begin{figure}[htbp]
    \centering
    \includegraphics[width=0.95\columnwidth]{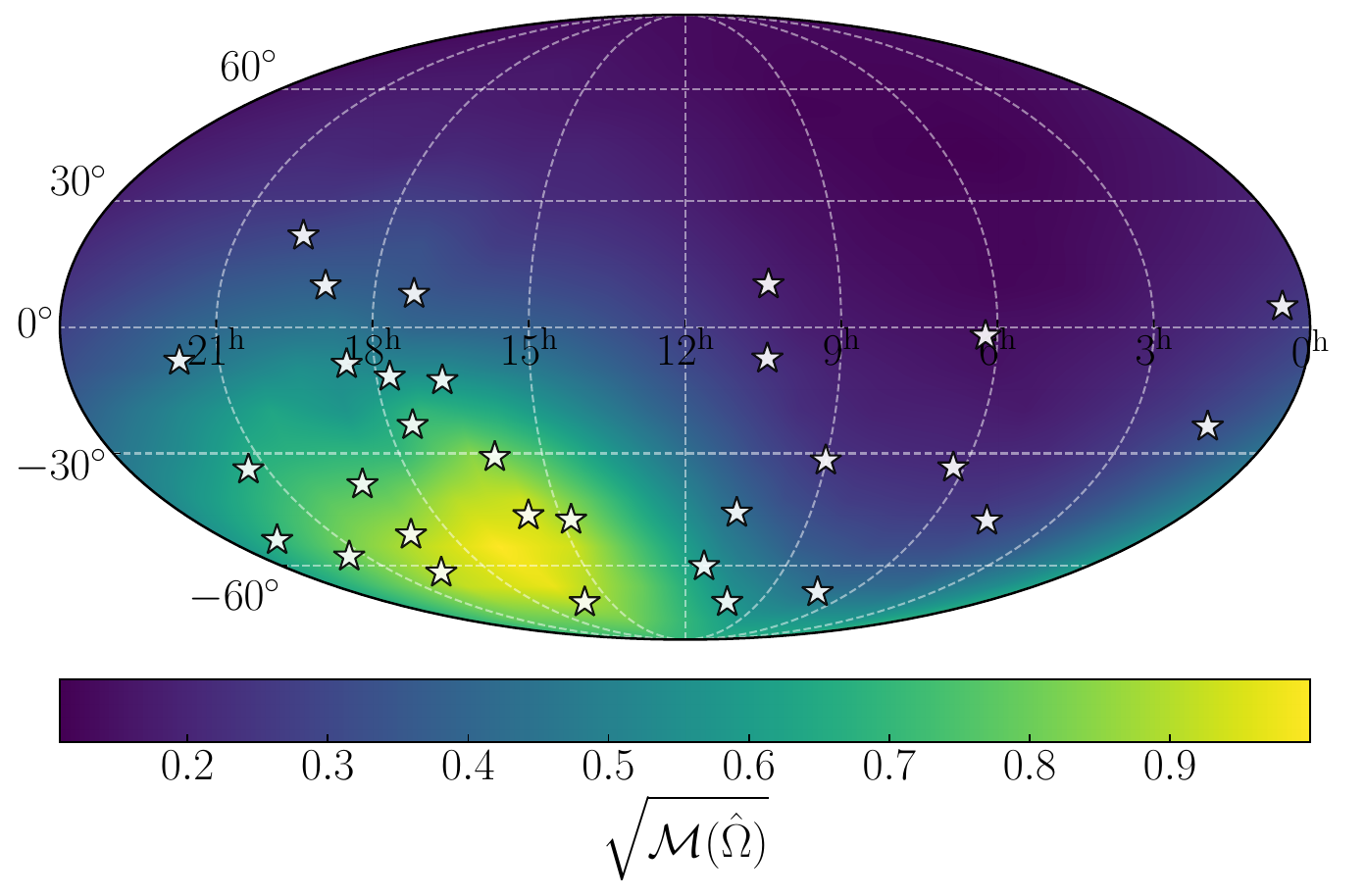}
    \caption{Sky map of the expected GW sensitivity for the PPTA DR3. The color map represents $\sqrt{\mathcal{M}(\hat{\Omega})}$, which quantifies the directional sensitivity to GWB anisotropies across the sky. The sensitivity distribution closely follows the spatial density of pulsars, with the highest sensitivity achieved in regions where the pulsar population is most concentrated.    
    } 
    \label{sensitivity}  
\end{figure}

\Fig{sensitivity} presents the sensitivity sky map constructed from PPTA DR3, where the sensitivity in each pixel is defined as the normalized square root of the corresponding diagonal element of the Fisher information matrix. The sensitivity distribution exhibits a clear correlation with the spatial distribution of pulsars across the sky. In particular, higher sensitivity is achieved in the high-latitude southern sky, where the PPTA provides denser coverage, resulting in a stronger directional response. This trend is consistent with findings reported by other PTA collaborations, such as NANOGrav and MPTA, in similar analyses. 

\section{Anisotropy search assuming a power-law GWB model}\label{power-law}

\begin{figure*}[t]
    \begin{center}
        \includegraphics[width=1\columnwidth]{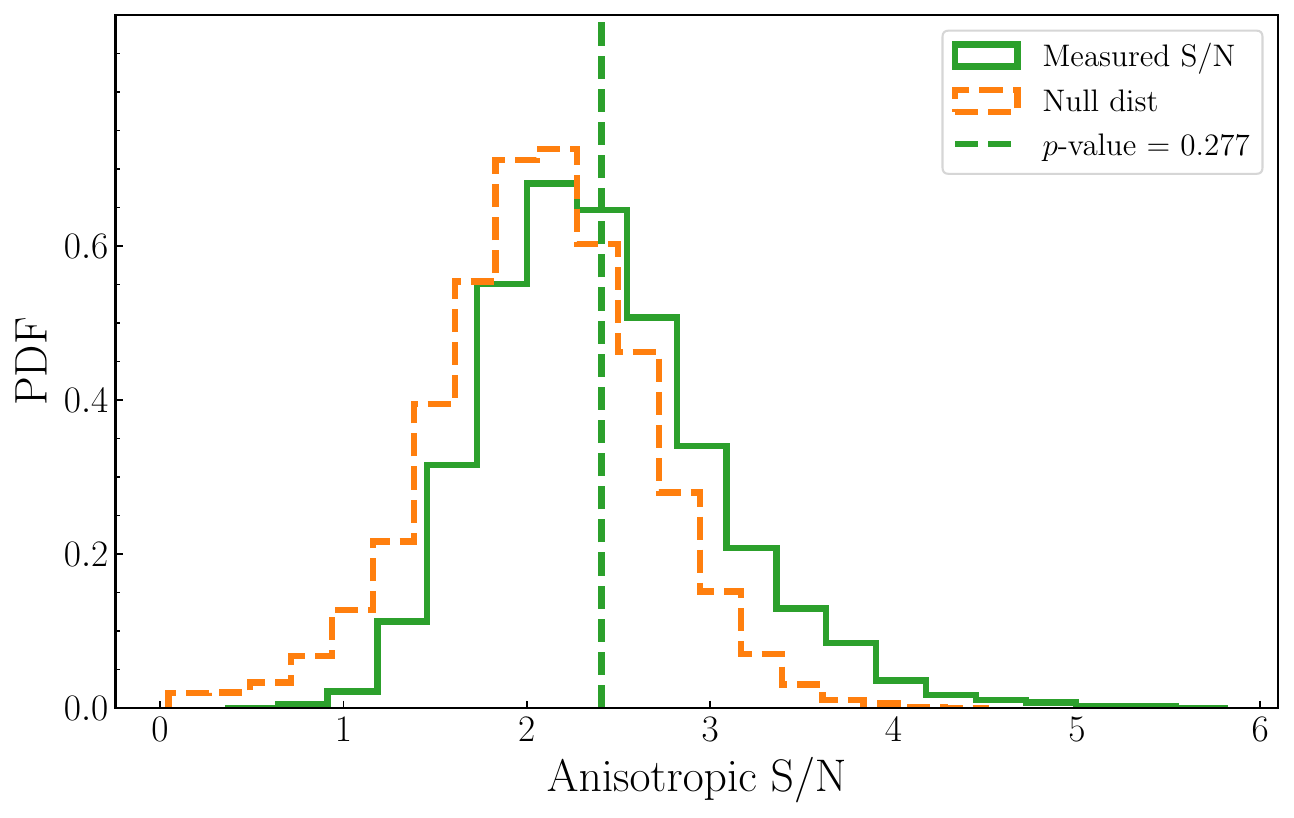}
        \includegraphics[width=1\columnwidth]{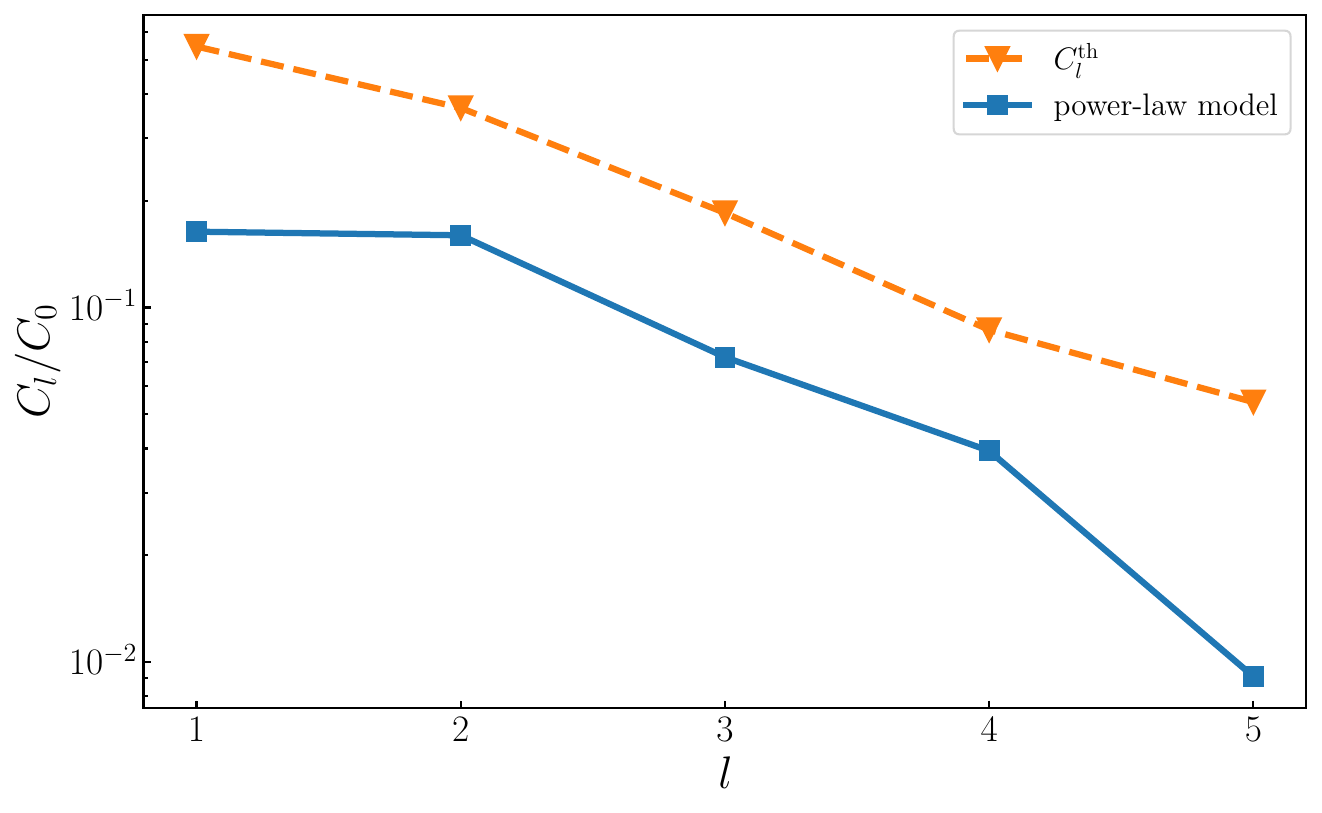}
        \caption{The results in both panels are obtained assuming a power-law strain spectrum.       
        \textbf{Left panel:} The noise-marginalized anisotropic SNR distribution (solid green line) and the SNR distribution under the null hypothesis (assuming an isotropic background; dashed orange line). The green vertical line marks the mean SNR, and the corresponding $p$-value from the null distribution is 0.277. \textbf{Right panel:} The blue solid line represents the measured angular power spectrum, while the orange dashed line denotes the detection threshold.
        } 
        \label{sqr_snr_Cl}  
    \end{center}
\end{figure*}

\begin{figure*}[t]
    \centering
    \includegraphics[width=0.45\textwidth]{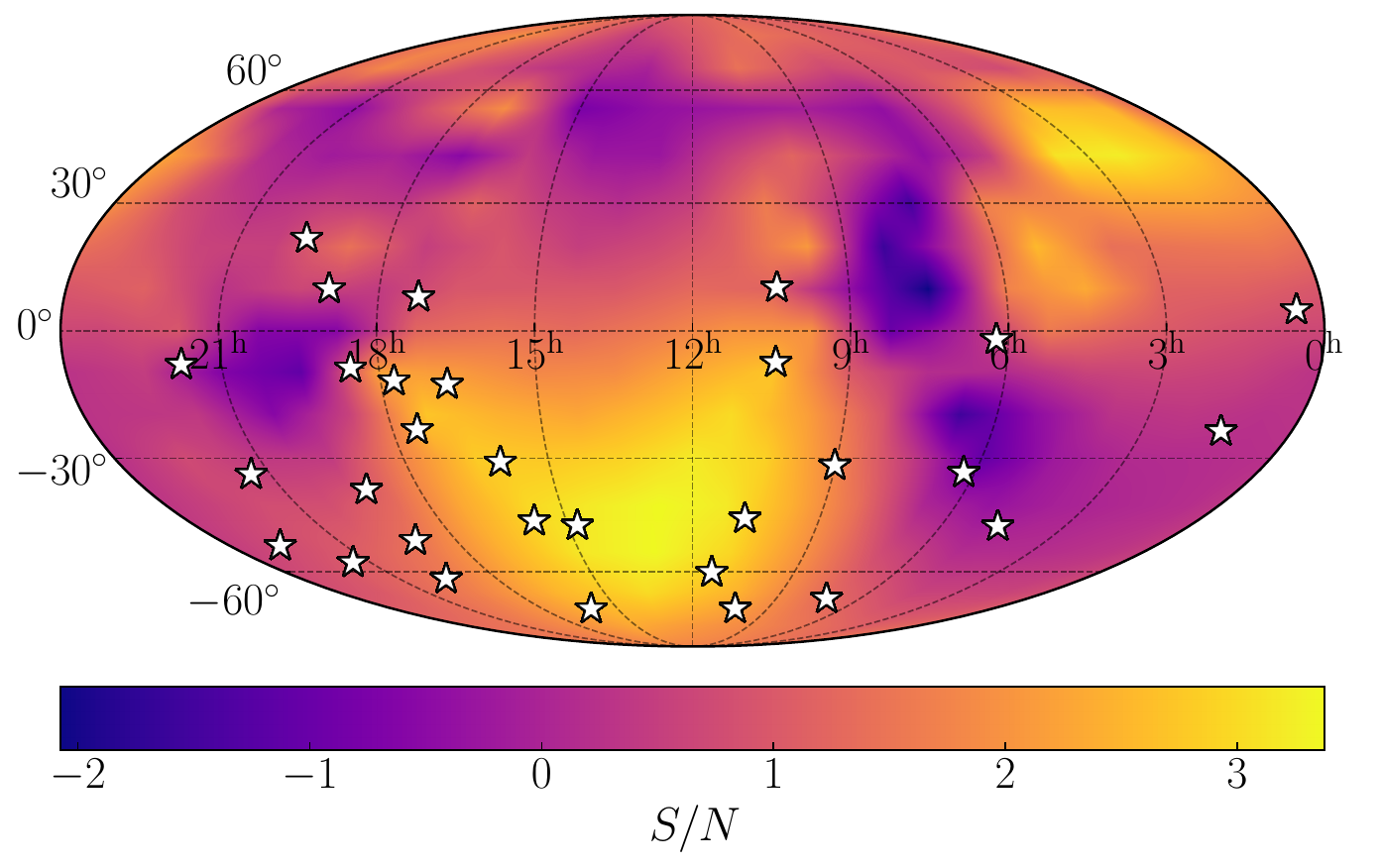}
    \includegraphics[width=0.45\textwidth]{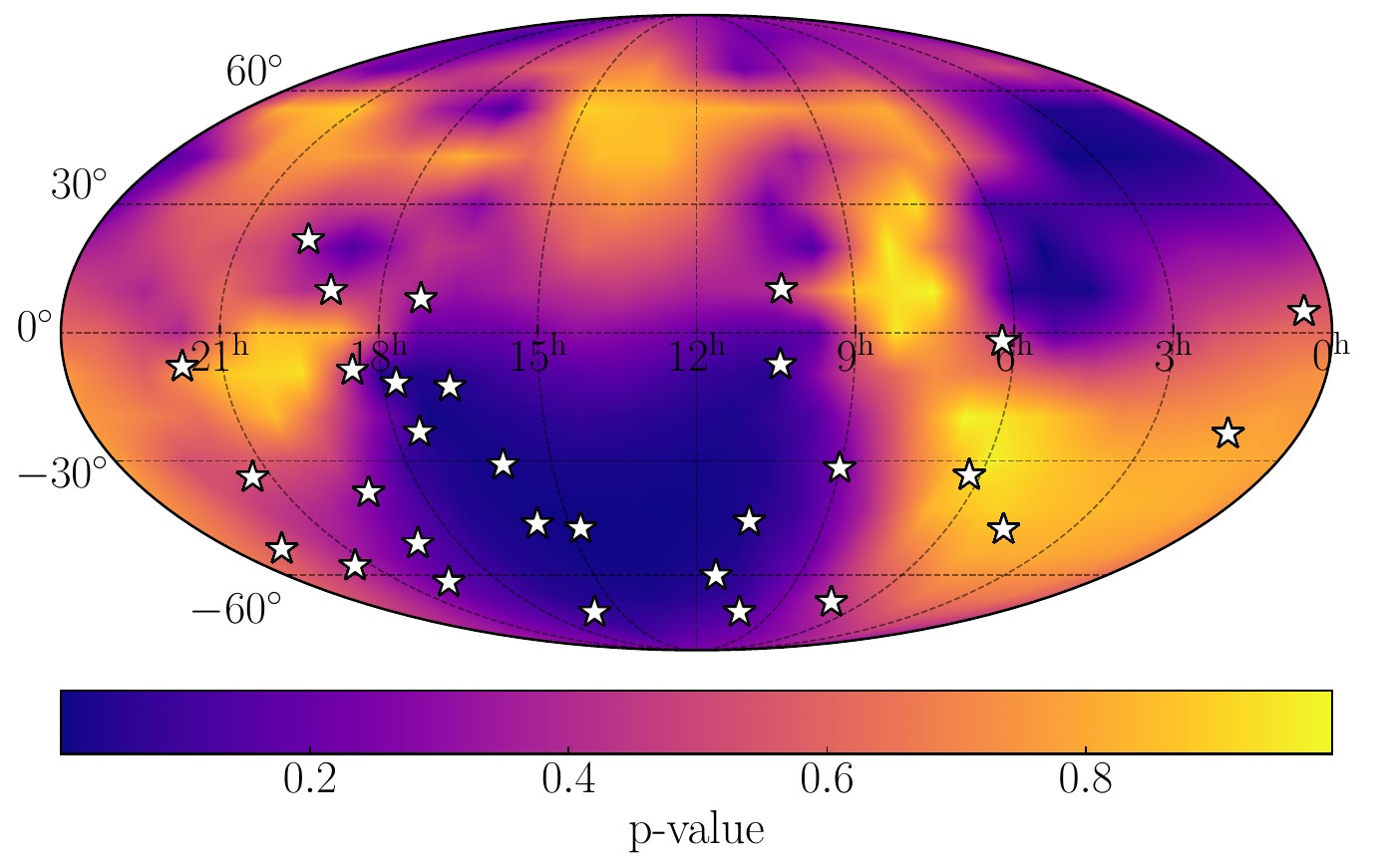}
    \caption{The sky maps are reconstructed assuming a power-law strain spectrum.
    The GWB sky map reconstructed based on the radiometer basis ($N_{\text{pix}}=192$). White stars mark the positions of the PPTA pulsars. The left panel shows the SNR in each pixel, while the right panel displays the corresponding $p$-value distribution. Although some directions exhibit relatively high SNR values, all $p$-values remain above the detection threshold of $3 \times 10^{-3}$, indicating consistency with an isotropic GWB.  
    }  
    \label{rdio}
\end{figure*}

Traditional PTA analyses often assume that the GWB can be described by a power-law spectrum, where the characteristic strain satisfies $h_c(f) \propto f^{-\gamma}$, with $\gamma = 2/3$ corresponding to the GWB produced by a population of inspiraling SMBBHs. In the main body of this study, we adopt a “free-spectrum” analysis approach, with no assumption about the spectral shape to enhance sensitivity to potential frequency-dependent structures. However, in order to further assess the impact of the power-law assumption on the results and make full use of the data across the entire frequency band, we supplement our analysis in this appendix with a search for anisotropy assuming a power-law GWB model.

Under the square-root spherical harmonic basis, the analysis results are presented in \Fig{sqr_snr_Cl}. The left panel shows the distribution of the measured anisotropic SNR, obtained via a noise-marginalization procedure. While Eq.~(\ref{eq:SNR}) defines the SNR for a fixed set of pulsar noise parameters, these parameters—such as the red-noise amplitudes and spectral indices—are characterized by significant posterior uncertainty. To account for this, we employ the noise-marginalized optimal statistic framework \cite{Vigeland_2018}. Specifically, we evaluate Eq.~(\ref{eq:SNR}) across a large number of pulsar-noise realizations sampled from their joint posterior distribution. The resulting histogram (solid green) thus represents the distribution of SNR values consistent with our noise characterization, with the vertical line denoting the mean. This is compared against the expected distribution under the null hypothesis (dashed orange), which is constructed from realizations of an isotropic GWB.
The mean measured SNR is approximately 2.4, corresponding to a $p$-value of 0.277. Since this value is much higher than the threshold of $3 \times 10^{-3}$, we cannot reject the null hypothesis of isotropy. The right panel shows the GWB angular power calculated from the observed data, and the detection threshold constructed based on the null hypothesis. Within the range of multipole moments analyzed, the measured angular power does not exceed the detection threshold and does not reach the $3\sigma$ significance level. Therefore, this analysis yields no statistically significant evidence for GWB anisotropy under the power-law model.

Fig. \ref{rdio} shows the reconstructed sky map using the radiometer pixel basis. According to the diffraction limit, in order to obtain an appropriate angular resolution, the analysis was performed on a HEALPix grid with $N_{\text{side}} = 4$. From the SNR map, it can be seen that the corresponding $p$-value of the maximum SNR is well above $3 \times 10^{-3}$. This indicates that, at the current sensitivity level, there is no statistically significant evidence for anisotropy in the GWB. It is worth noting that the $p$-values shown here are calculated differently from those presented in the main text. In this appendix, for each realization under the null hypothesis, we compute the detection statistic independently for every pixel in the reconstructed sky map. This allows us to obtain the null distribution and corresponding $p$-value for each pixel. If the $p$-value is smaller than $3 \times 10^{-3} / N_{\mathrm{pix}}$, the reconstructed sky map is considered globally significant at approximately $3\sigma$. Here, the trial factor $1/N_{\mathrm{pix}}$ serves as a correction, converting the local significance of individual pixels into a global significance measure.

\bibliographystyle{apsrev4-1}
\bibliography{ref}

\end{document}